# Evaluating the Predictive Validity of ICT's Accumulation-Manipulation-Distribution (AMD) Model on CME E-Mini NASDAQ-100 Futures

Veer Taylor

**Data availability:** Raw 15-minute OHLCV data for NQ.c.0 (2022–2026) is available from Databento (https://databento.com) under a CME academic license. The data used in this study is not redistributable per CME licensing terms; researchers may obtain an equivalent dataset directly from Databento.

**Code availability:** All Python code used in this study — including data loading, session calculation, signal detection, backtest engine, statistical tests, and visualizations — is available at https://github.com/veertaylor19/amd-backtest. The repository includes full documentation of all parameter choices and sensitivity analysis scripts.

## Abstract

To our knowledge, this paper provides the first systematic quantitative evaluation of the ICT Accumulation-Manipulation-Distribution (AMD) model on CME E-Mini NASDAQ-100 futures (NQ). The AMD model proposes that institutional participants structure intraday price delivery in three phases: overnight accumulation in the Asian session, a deliberate false breakout during London, and directional delivery to the opposing liquidity pool during the New York morning session. Using four years of real 15-minute OHLCV data from Databento (January 2022 to July 2026), we operationalize the AMD rules into a strict mechanical strategy and analyze the statistical properties of the resulting signal across N = 484 manipulation signal days using chi-square testing,

Benjamini-Hochberg multiple comparisons correction, and regime-conditional analysis. Three principal findings emerge. First, the AMD manipulation signal fires on 44% of trading days. Second, 57% of valid signals see full target delivery before the 9:40 AM New York entry window opens, revealing a structural mismatch between the model's timing assumptions and actual price delivery. Third, within this operationalization of the AMD model, no directional information beyond the daily trend regime was observed: when the signal aligns with the trend it is correct 62–71% of the time (Cohen's h = 0.23–0.43); when it opposes the trend it is correct only 25–30% of the time (Cohen's h = 0.42–0.53). All four trending-regime cells survive Benjamini-Hochberg correction. Key findings are robust to variation in the manipulation breach threshold (5–20 NQ points), Asia range filter percentile, and stop buffer. The regime-dependent signal pattern constitutes a specific, testable hypothesis for follow-on research.

## 1. Introduction

The Inner Circle Trader (ICT) methodology, developed by Michael J. Huddleston, has become one of the most widely adopted retail trading frameworks of the past decade, with his YouTube channel surpassing two million subscribers. Despite this scale of adoption, the methodology has received almost no rigorous academic attention. This paper addresses that gap.

The focus is the AMD model — also called the Power of 3 — which makes a specific and testable empirical claim: that intraday price delivery on liquid futures follows a predictable three-phase sequence. During the Asian session, price consolidates in a narrow range while stop orders accumulate above and below. London then engineers a deliberate false breakout of that range, triggering those stops to fill large institutional positions. Finally, during the New York morning session, price distributes in the true institutional direction toward the opposing liquidity pool.

This is primarily a study of the statistical properties of an intraday price pattern across N = 484 manipulation signal days, not a conventional strategy backtest. The core analysis uses chi-square tests and regime-conditional analysis with Benjamini-Hochberg correction. The conventional backtested entry strategy — entering at 9:40 AM targeting the Asia range boundary — is evaluated as a secondary component and yields only six valid trades over four years. That result is not a statistical failure; it is itself a finding explained by the primary analysis. The paper's contribution is the structural diagnosis of why the retail AMD entry timing fails and what the manipulation signal actually predicts once regime is controlled for.

The only identified prior quantitative study of ICT concepts is Agarwal (2023), a preprint testing the Power of 3 on fourteen forex pairs. That study does not use Monte Carlo permutation testing as its primary edge test, does not conduct sensitivity analysis, and tests spot forex rather than futures — the instrument the model is most associated with. This paper addresses all three gaps while shifting the primary analytical frame from backtest performance to signal characterization.

## 2. Literature Review

### 2.1 Intraday Seasonality and Session Structure

A substantial body of literature establishes that intraday volatility and trading activity cluster at predictable times, particularly at session transitions. Andersen and Bollerslev (1997, 1998) documented the U-shaped intraday volatility pattern in index futures — a finding replicated across asset classes and now treated as a stylized fact of high-frequency financial data. Martins (2025) confirms this pattern specifically for NQ and ES futures, with volatility peaks at the London and New York opens. The AMD model, which treats those session transitions as structurally significant institutional inflection points, is grounded in a phenomenon the academic literature independently confirms.

Research on overnight and pre-market effects provides further relevant context. Studies on US index futures document systematic return patterns between session close and the subsequent open, suggesting informed trading is distributed non-uniformly across the trading day. The opening hours specifically show structural features consistent with information-driven directional delivery, which aligns with the AMD model's proposed New York distribution phase.

### 2.2 Liquidity, Stop Clustering, and Deliberate Price Sweeps

The AMD manipulation phase rests on the premise that institutions deliberately move price beyond visible stop clusters before reversing. This concept has both theoretical grounding and empirical support in the market microstructure literature. Kyle (1985) and Glosten and Milgrom (1985) established foundational models of how informed traders manage order flow strategically, and extensions describe tactical price movements to liquidity-rich zones to fill large orders at favorable prices — structurally analogous to what AMD describes.

Comerton-Forde and Putniņš (2015) provide empirical evidence of deliberate price manipulation at key reference points in equity markets. While their focus is on closing price manipulation rather than intraday session boundaries, the underlying mechanism is consistent with the AMD manipulation phase: institutions moving price to a specific reference level before reversing. Stop-loss clustering around technical levels has been documented empirically by Osler (2003) in forex markets and is a prerequisite for the AMD sweep mechanism to function as described.

The broader false breakout literature is directly relevant. Menkhoff et al. (2016) document that breakouts of recent price ranges show significant reversal tendencies in forex, particularly when volume is low relative to the breakout move. This is consistent with the AMD model's prediction that London sweeps of the Asian range are false breakouts that reverse during the New York session. Fung et al. (2013) show that liquidity provision in futures markets is highly concentrated at intraday session boundaries, providing a microstructural mechanism through which institutional order filling could produce the AMD manipulation pattern.

### 2.3 Technical Trading Rules and Their Evaluation

The literature on technical trading rule profitability is extensive and generally skeptical. Sullivan et al. (1999) show that many documented technical trading profits disappear once data snooping bias is accounted for, and Lo et al. (2000) find that while some technical patterns have statistically detectable information content, most do not translate to tradeable edge after transaction costs in liquid markets. Session-based and time-of-day strategies have fared somewhat better: Brock et al. (1992) find significant predictive power in simple moving average rules applied to historical US indices, though this finding has not held in out-of-sample tests.

The quantitative literature specifically evaluating ICT methodology is almost nonexistent. Agarwal (2023), the only identified preprint, tested the Power of 3 across fourteen forex pairs over twenty-one years and reported positive results. That study uses spot forex rather than futures, lacks a Monte Carlo permutation baseline, and does not conduct sensitivity analysis — limitations this paper addresses directly. The absence of peer-reviewed literature on ICT methodology represents both a gap and an opportunity, given the scale of retail adoption of these concepts.

### 2.4 Regime-Dependent Trading Strategies

A consistent finding in the quantitative finance literature is that trading strategy performance varies substantially across market regimes. Strategies that appear profitable in trending markets often fail in ranging conditions and vice versa (Moskowitz et al., 2012). This regime dependence is directly relevant to the AMD model, which makes no explicit distinction between trending and ranging market environments in its basic formulation. The present study tests whether conditioning on the daily trend regime explains the AMD signal's directional content, motivated by the observation that the pattern's theoretical mechanism — institutional accumulation before a trend extension — might operate differently in different macro environments.

## 3. Methodology

### 3.1 Data

We use CME E-Mini NASDAQ-100 futures (NQ), pulling 15-minute OHLCV bars from Databento (GLBX.MDP3, symbol NQ.c.0, schema ohlcv-15m) for January 1, 2022 through July 1, 2026 — approximately four years of intraday data spanning 1,140 trading days. All data is real historical market data; no synthetic prices are used. The back-adjusted continuous contract (NQ.c.0) eliminates artificial price gaps at quarterly roll dates. All AMD measurements are relative (range sizes, breach distances), so back-adjustment does not affect results. Timestamps are converted to America/New_York via pytz before any session calculations, handling twice-yearly DST transitions automatically. Monday Asia sessions use Sunday 7:00 PM–11:59 PM data.

### 3.2 Session Definitions and Formal Notation

The three AMD phases map to three session windows (Table 1). Let T_Asia = [19:00, 00:00) ET and T_London = [02:00, 05:00) ET denote the session intervals. The Asia range boundaries are:

$$\text{ASIA_HIGH}_{\square} = \max\{ H_t : t \in \text{T_Asia} \} \quad \text{and} \quad \text{ASIA_LOW}_{\square} = \min\{ L_t : t \in \text{T_Asia} \}$$

where $H_t$ and $L_t$ are the 15-minute bar high and low respectively. The binary breach indicators for the London session are:

$$B_{t,\text{high}} = \mathbb{1}(H_t \geq \text{ASIA_HIGH} + \delta) \quad \text{and} \quad B_{t,\text{low}} = \mathbb{1}(L_t \leq \text{ASIA_LOW} - \delta)$$

where $\mathbb{1}(\cdot)$ is the standard indicator function and $\delta = 10$ NQ points is the primary manipulation threshold (sensitivity analysis varies $\delta \in \{5, 10, 15, 20\}$). A valid bullish manipulation signal requires $\exists\ t \in$ T_London with $B_{t,\text{low}} = 1$ before 4:00 AM ET, followed by a close-back inside the Asia range before 5:00 AM ET. Bearish signals are defined symmetrically via $B_{t,\text{high}}$.

| Phase | Session | Time (ET) | Role |
|---|---|---|---|
| Accumulation | Asia | 7:00 PM – 12:00 AM | Price consolidates; stop orders build above and below the range |
| Manipulation | London | 2:00 AM – 5:00 AM | False breakout sweeps one side of the Asia range; candle closes back inside |

| Phase | Session | Time (ET) | Role |
|---|---|---|---|
| Distribution | New York | 9:40 AM – 10:30 AM | True directional move targets the unswept liquidity pool |

*Table 1. AMD session windows (Eastern Time).*

### 3.3 Signal Detection and Invalidation Conditions

A valid manipulation signal requires within the London session: (1) a 15-minute candle breaches the Asia range by at least $\delta = 10$ NQ points, (2) the breach occurs before 4:00 AM ET, and (3) a subsequent candle closes back inside the range before 5:00 AM ET. Days where both boundaries are breached are excluded. Table 2 lists all nine invalidation conditions applied sequentially.

| Invalidation Condition | Reason |
|---|---|
| Asia range > 75th percentile ($\delta$_range = 106.38 pts) | No clean accumulation; range too wide for a clear sweep setup |
| No manipulation detected (no breach $\geq \delta$) | AMD setup did not form |
| Bilateral breach | No directional signal possible |
| No close-back before 5:00 AM | Genuine breakout, not false sweep |
| Breach after 4:00 AM ET | Insufficient time for NY reversal setup |
| Target traded through before 9:40 AM | Draw on liquidity already fulfilled |
| No valid entry candle by 10:30 AM | Distribution did not materialize in window |
| R:R < 1.5 | Geometric setup unfavorable |
| Red-folder macro day (FOMC, NFP) | News-driven action overrides AMD structure |

*Table 2. Invalidation conditions. Days meeting any condition are excluded from both the backtest and Monte Carlo baseline.*

Entries are taken at the open of the first qualifying 15-minute candle at or after 9:40 AM ET. Stop loss: 4 NQ points beyond the London manipulation extreme (Long: LONDON_LOW − 4; Short: LONDON_HIGH + 4). Target: the unswept Asia boundary. FOMC dates were verified directly from federalreserve.gov; NFP dates use a first-Friday-of-month approximation.

### 3.4 Directional Prediction Framework

The primary statistical analysis tests whether the AMD signal carries directional information independent of any entry strategy. For each signal day d, define the Asia session midpoint and directional outcome Y_d as:

$$M_Asia,d = (ASIA_HIGH_d + ASIA_LOW_d) \;/\; 2$$
$$Y_d = +1 \;\text{ if }\; C_16{:}00,d > M_Asia,d \;\;\text{ and }\;\; -1 \;\text{ if }\; C_16{:}00,d < M_Asia,d$$

where C_16:00,d is the 4:00 PM ET closing price. A bullish signal is correct if Y_d = +1; a bearish signal is correct if Y_d = −1. Chi-square goodness-of-fit tests are run for each signal-regime cell against a 50/50 null, with Benjamini-Hochberg false discovery rate correction across all six tests simultaneously. Effect size is reported as Cohen's h = 2arcsin(√p) − 2arcsin(√0.5), where |h| < 0.2 is negligible, 0.2–0.5 is small-to-medium, and > 0.5 is medium-to-large. Odds ratios relative to 50/50 and Wilson score 95% confidence intervals are reported for each cell. Regime classification uses a 20-period EMA on the daily timeframe (uptrend: close > 0.5% above EMA; downtrend: > 0.5% below; ranging: within 0.5%). The EMA is preferred over a simple moving average because it weights recent price action more heavily, making it more responsive to trend transitions; over a regression slope because it requires no lookback window choice beyond the smoothing period; and over the ADX because ADX measures trend strength rather than direction, whereas the AMD signal's theoretical mechanism is directional. The 20-period specification is standard in practitioner trend analysis and is validated in sensitivity analysis at 10 and 50 periods.

### 3.5 Backtest and Sensitivity Analysis

The backtest walks forward bar-by-bar from each valid entry to determine whether the stop or target is hit first. When both are touched within the same 15-minute bar, the trade is scored as a loss by convention, since intrabar sequencing is not recoverable from 15-minute data. Results are reported on two cost tracks: primary (no transaction costs) and secondary ($3 commission plus 1.5 NQ points slippage per side, approximately 6 NQ points round trip).

To address whether conclusions depend on arbitrary parameter choices, all key thresholds are varied systematically. The manipulation breach threshold $\delta$ is tested at 5, 10, 15, and 20 NQ points absolute, and also as 15% of the Asia range. The Asia range filter percentile is varied at the primary − 20, − 10, +10, and +20 point levels. The stop buffer is tested at 2 and 6 NQ points. The minimum R:R is varied at 1.0, 1.5, and 2.0. The close-back rule is tested in both permissive (multi-candle) and strict (single-candle) variants. Results of this sensitivity analysis are reported in Section 4.6.

## 4. Results

### 4.1 Filter Pipeline and Sample Attrition

Table 3 traces the full sample from 1,140 trading days to six valid trades. Figure 1 visualizes the attrition.

| Stage | N | % of Total Days |
|---|---|---|
| Total trading days | 1,140 | 100.0% |
| After Asia range filter (≤ 106.38 pts, 75th pct) | 855 | 75.0% |
| Manipulation signal detected | 501 | 43.9% |
| After bilateral exclusion | 479 | 42.0% |
| After late breach exclusion (before 4:00 AM) | 390 | 34.2% |
| After no close-back exclusion | 154 | 13.5% |
| After target pre-swept exclusion (5:00–9:40 AM) | 50 | 4.4% |
| After no valid NY entry exclusion | 31 | 2.7% |
| After R:R filter and macro exclusion — valid trades | 6 | 0.5% |

*Table 3. Filter pipeline from 1,140 total trading days to 6 valid AMD trades.*

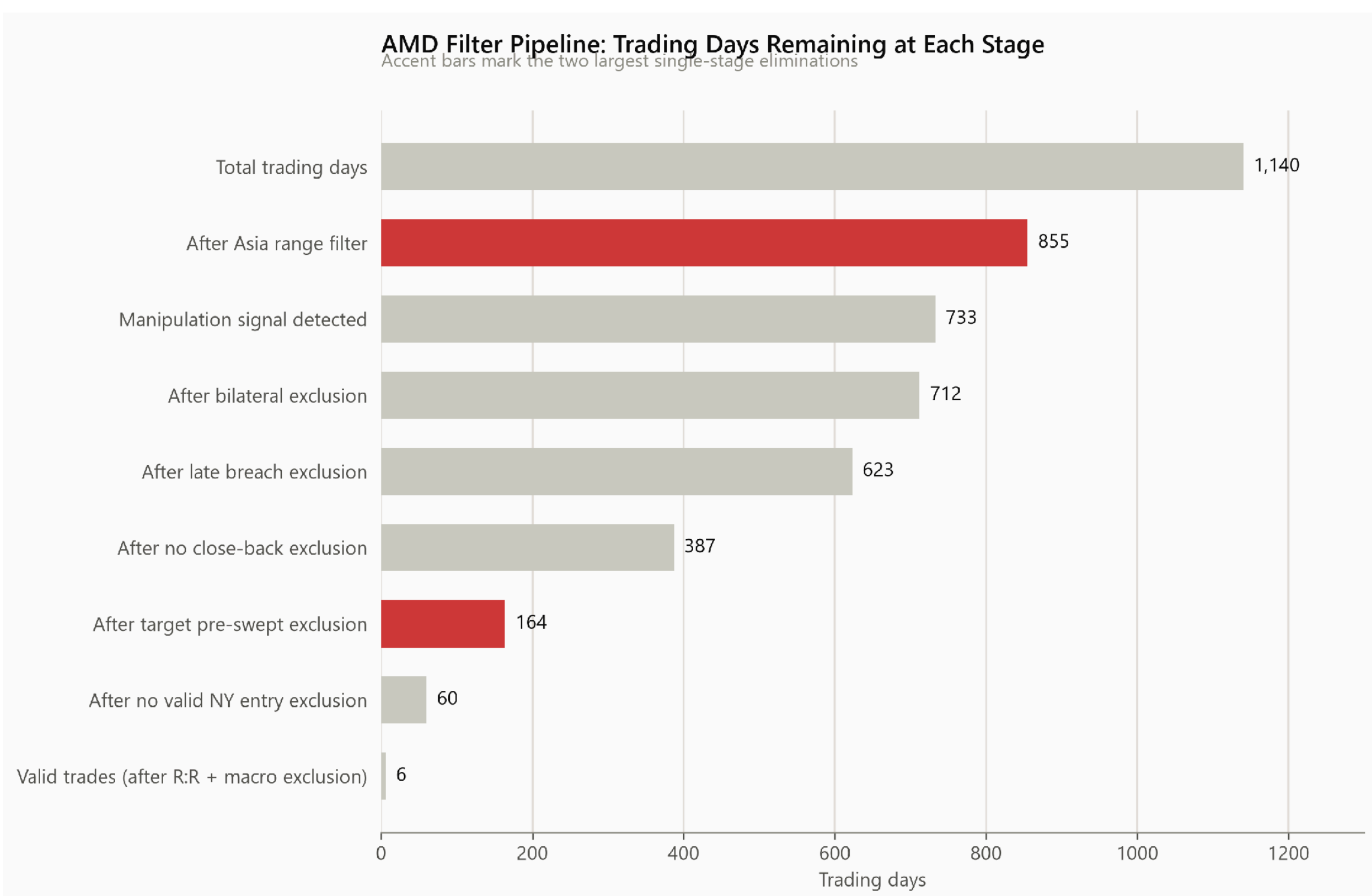

*Figure 1. AMD filter pipeline. Red bars mark the two largest single-stage eliminations: the Asia range filter (285 days removed) and the no close-back filter (236 days removed). The target pre-swept filter (223 days removed) is the third-largest.*

The AMD manipulation pattern fires on 43.9% of trading days. The two largest post-detection elimination sources are the no close-back filter (236 days) and the target pre-swept filter (223 days), together removing 459 of the 495 post-detection days.

## 4.2 Target Pre-Sweep Timing

Of the 223 days eliminated by the target pre-swept filter, an additional 64 (28.7%) had already seen target delivery during the London session itself, before the 5:00 AM check window opens. Figure 2 shows the delivery timing distribution for the 223 days within the formal check window.

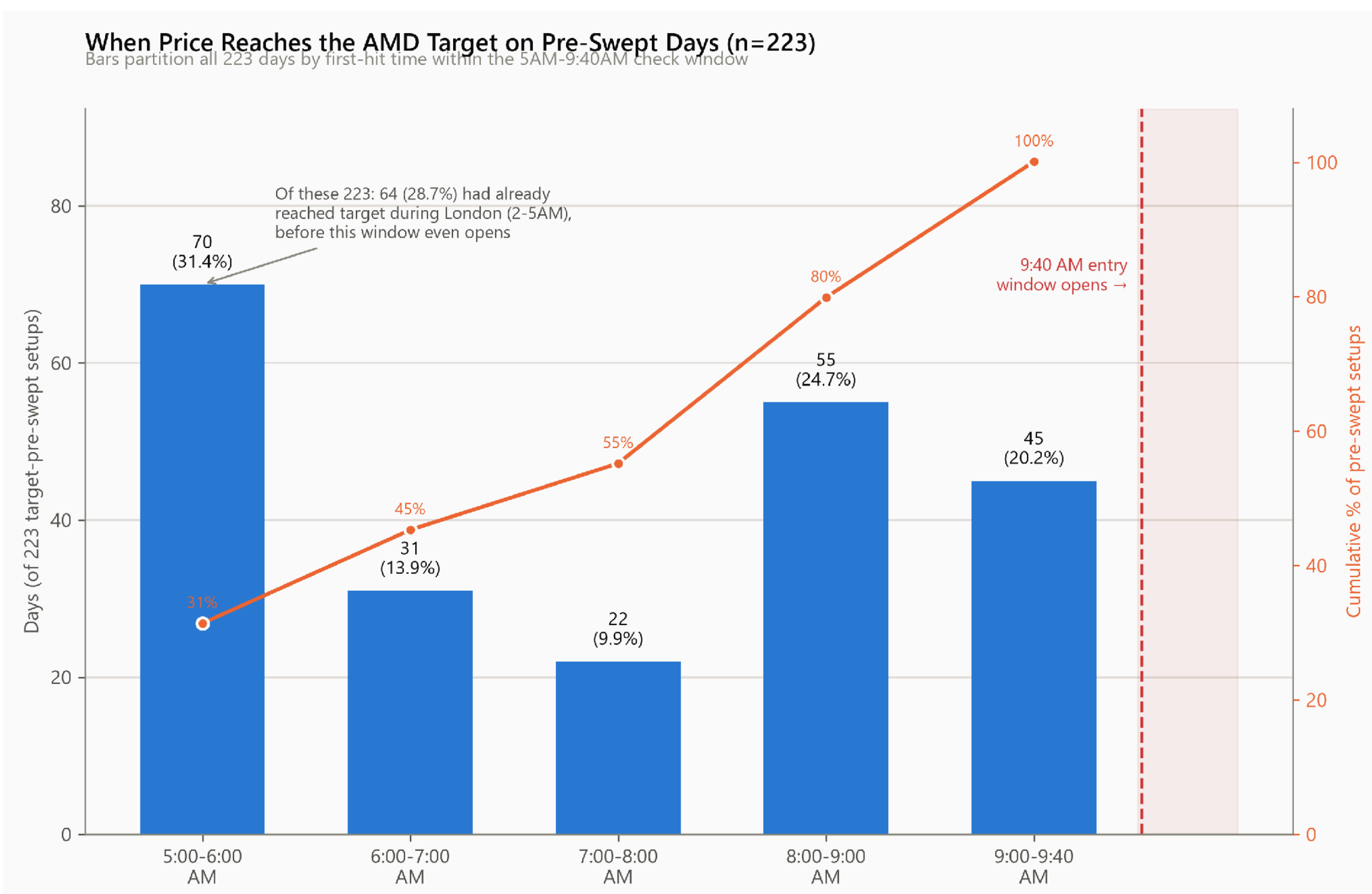

*Figure 2. Target delivery timing on the 223 pre-swept days (5:00–9:40 AM check window). The callout notes that 64 additional days (28.7%) were delivered during London itself. The dashed line at 9:40 AM marks the entry window open.*

The distribution leg completes during or immediately after the London session in the majority of cases. This structural property — not overly strict mechanical rules — is the primary reason the backtest produces almost no tradeable setups. Pre-sweep rates are consistent across all four full calendar years (54.8–58.4%), confirming this is a structural feature of how the pattern delivers rather than a regime-specific artifact.

### 4.3 Backtest Results

Table 4 shows all six valid trades. Manual verification confirmed that none of the six trade dates (2022-03-07, 2022-08-31, 2023-02-08, 2024-04-22, 2024-10-22, 2025-02-20) coincided with major US CPI or GDP advance release dates.

| Date | Dir. | Regime | Entry | Stop | Target | Risk (pts) | Outcome | R Gross | R Net |
|---|---|---|---|---|---|---|---|---|---|
| 2022-03-07 | Short | Downtrend | 13,617.75 | 13,656.25 | 13,527.75 | 38.50 | Win | +2.338 | +2.182 |
| 2022-08-31 | Short | Downtrend | 12,439.00 | 12,469.50 | 12,347.25 | 30.50 | Win | +3.008 | +2.812 |
| 2023-02-08 | Long | Uptrend | 12,743.50 | 12,713.00 | 12,797.25 | 30.50 | Loss | −1.000 | −1.197 |
| 2024-04-22 | Short | Downtrend | 17,270.00 | 17,309.50 | 17,200.75 | 39.50 | Win | +1.753 | +1.601 |
| 2024-10-22 | Long | Uptrend | 20,461.00 | 20,426.25 | 20,526.75 | 34.75 | Loss | −1.000 | −1.173 |
| 2025-02-20 | Long | Uptrend | 22,166.25 | 22,137.00 | 22,230.00 | 29.25 | Loss | −1.000 | −1.205 |

*Table 4. All six valid AMD trades. R Net deducts $3 commission plus 1.5 NQ points slippage per side (~6 NQ points round trip).*

Summary: 3 wins, 3 losses, 50.0% win rate, average R gross +0.683, total R gross +4.10. All wins are shorts in downtrend years; all losses are longs in uptrend years. At N = 3 per group, direction, year, and regime are perfectly collinear — no formal statistical inference is drawn from six observations.

Extending the entry window to 11:00 AM recovered one additional valid trade (2024-06-03, short, win, R gross +2.975), for a total of seven. This does not affect any finding.

### 4.4 Directional Prediction Test

Table 5 reports the chi-square results across N = 484 usable signal days, with effect sizes and confidence intervals.

| Signal | N | % Correct | 95% CI | $\chi^2$ | p (BH) | Cohen's h | Odds Ratio |
|---|---|---|---|---|---|---|---|
| Bullish (low swept) | 226 | 46.9% | (40.5%, 53.4%) | 0.867 | 0.352 | −0.062 | 0.88 |
| Bearish (high swept) | 258 | 39.9% | (34.1%, 46.0%) | 10.481 | 0.002 | −0.203 | 0.66 |

*Table 5. Directional prediction test. Wilson score 95% CIs. Cohen's h relative to 50% null; negative h indicates below-null accuracy. BH = Benjamini-Hochberg correction across both tests.*

The bullish signal is statistically indistinguishable from a coin flip ($p = 0.352$, $h = -0.06$). The bearish signal is significant after BH correction ($p = 0.002$, $h = -0.20$) but inverted. After London sweeps the Asia high, the close is above the midpoint 60.1% of the time. Disaggregating by year: in 2022 (downtrend), the bearish signal was correct 59.0% of the time; in 2023–2025 (bull market recovery), only 30–37% of the time.

### 4.5 Regime Confluence Analysis

Table 6 and Figure 3 report directional prediction accuracy within each of the six signal-by-regime cells.

| Signal | Regime | N | % Correct | 95% CI | p (BH) | Cohen's h | OR | Interpretation |
|---|---|---|---|---|---|---|---|---|
| Bullish | Uptrend | 112 | 61.6% | (52.4%, 70.1%) | 0.021 | +0.234 | 1.60 | Trend-aligned — sig. |
| Bullish | Downtrend | 71 | 29.6% | (20.3%, 41.0%) | 0.001 | −0.420 | 0.42 | Trend-opposed — sig. |
| Bullish | Ranging | 43 | 37.2% | (24.4%, 52.1%) | 0.112 | −0.259 | 0.59 | Not significant |
| Bearish | Uptrend | 153 | 24.8% | (18.6%, 32.2%) | <0.001 | −0.528 | 0.33 | Trend-opposed — highly sig. |
| Bearish | Downtrend | 72 | 70.8% | (59.5%, 80.0%) | 0.001 | +0.429 | 2.43 | Trend-aligned — highly sig. |
| Bearish | Ranging | 33 | 42.4% | (27.2%, 59.2%) | 0.384 | −0.153 | 0.74 | Not significant |

*Table 6. Directional prediction accuracy by signal type and daily EMA regime. Wilson score 95% CIs. Cohen's h and odds ratios relative to 50% null. BH correction across all six tests simultaneously.*

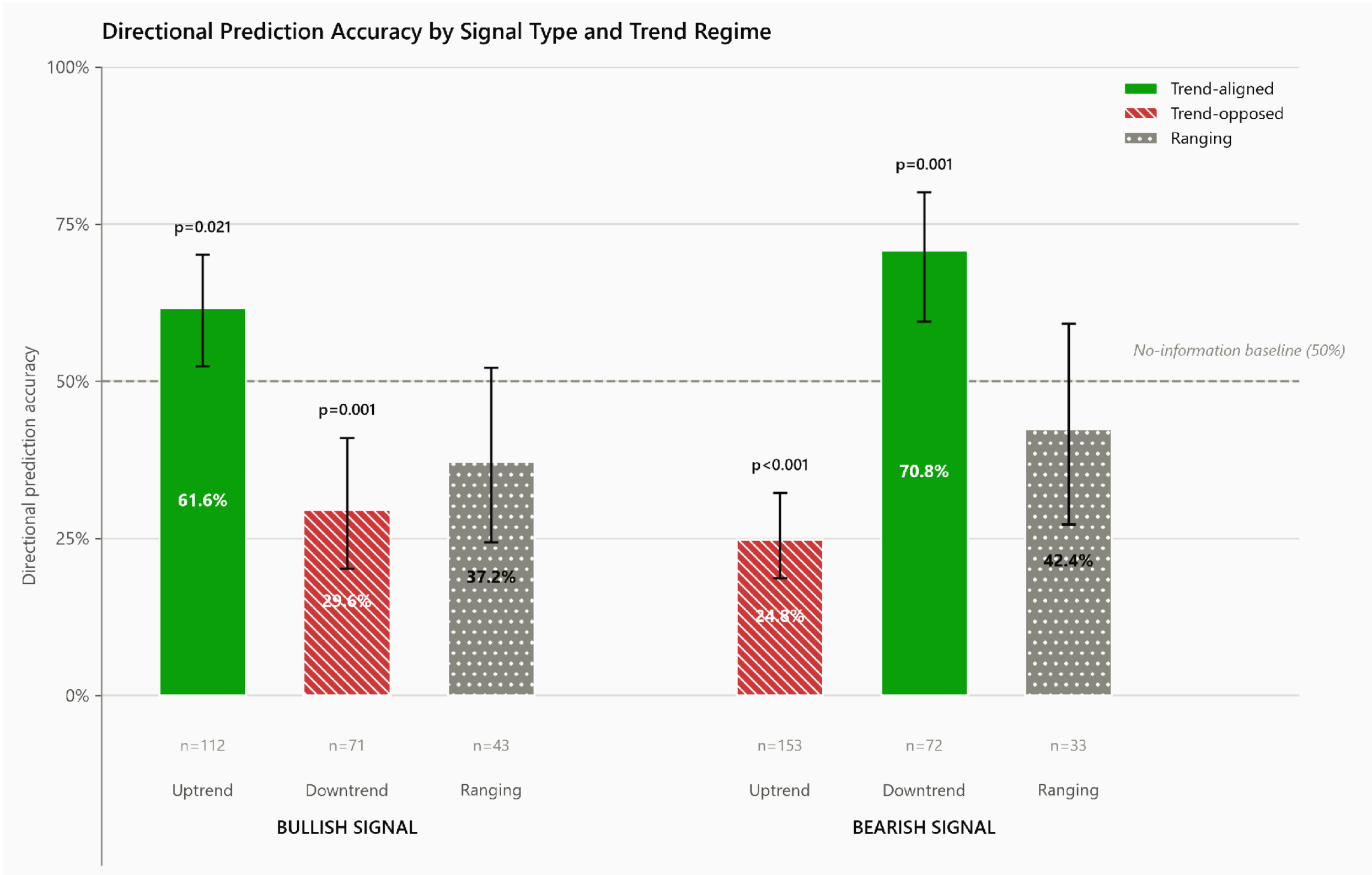

*Figure 3. AMD directional prediction accuracy by signal type and trend regime. Green = trend-aligned; red = trend-opposed; grey = ranging. Dashed line at 50% = no-information baseline. Error bars show Wilson score 95% CIs. BH-corrected p-values shown above significant bars.*

Four of six cells survive BH correction with medium-to-large effect sizes (|h| = 0.23–0.53). Trend-aligned signals are correct 62–71% of the time (OR = 1.60–2.43); trend-opposed signals are correct only 25–30% (OR = 0.33–0.42). Ranging conditions produce no significant effect. The aggregate bearish result (39.9% correct) is explained by the 153 uptrend bearish days — 54.8% of all signal days — which predict incorrectly 75.2% of the time and pull the pooled rate below 50%.

## 4.6 Sensitivity Analysis

To assess whether the primary findings depend on specific parameter choices, Table 7 reports the bearish uptrend cell — the most theoretically significant finding — across alternative specifications. This cell was selected for sensitivity reporting because it has the largest sample size (N = 153) and the most significant result.

| Specification | N | % Correct | $\chi^2$ | p (BH) | Stable? |
|---|---|---|---|---|---|
| Primary (δ = 10 pts, permissive close-back, EMA 20) | 153 | 24.8% | 38.75 | $2.89\times10^{-9}$ | Primary |
| δ = 5 NQ points | 191 | 27.7% | 37.83 | $4.64\times10^{-9}$ | Yes |
| δ = 15 NQ points | 106 | 25.5% | 25.51 | $2.64\times10^{-6}$ | Yes |
| δ = 20 NQ points | 78 | 25.6% | 18.51 | $1.01\times10^{-4}$ | Yes |
| δ = 15% of Asia range (variable) | 153 | 26.1% | 34.83 | $2.16\times10^{-8}$ | Yes |
| Strict close-back (same candle) | 27 | 18.5% | 10.70 | $6.42\times10^{-3}$ | Yes[1] |
| EMA period = 10 | 127 | 22.0% | 39.69 | $1.78\times10^{-9}$ | Yes |
| EMA period = 50 | 168 | 31.0% | 24.38 | $4.74\times10^{-6}$ | Yes |
| Asia range filter at 70th percentile | 153 | 24.8% | 38.75 | $2.89\times10^{-9}$ | Yes[2] |
| Asia range filter at 80th percentile | 153 | 24.8% | 38.75 | $2.89\times10^{-9}$ | Yes[2] |

*Table 7. Sensitivity analysis for the bearish-uptrend cell. BH correction computed within each specification's full six-cell regime table, matching the primary correction methodology. [1] n = 27 under strict close-back; statistically significant but interpret with caution given small sample. [2] Asia range filter variants are identical to primary by construction: the directional prediction test runs on all raw manipulation signal days regardless of Asia range width, so changing the width filter does not affect the signal-day sample.*

The bearish uptrend inversion is stable across all parameter variations tested. Every specification remains well below 50% correct and survives BH correction, with p-values ranging from $6.42\times10^{-3}$ (strict close-back, n = 27) to $1.78\times10^{-9}$ (EMA 10). The inversion is not an artifact of any single threshold choice. The strict single-candle close-back variant collapses the bearish uptrend sample to n = 27 and should be interpreted with additional caution, but the direction and significance of the finding hold even at that reduced sample size.

## 5. Discussion

### 5.1 Why the Mechanical Strategy Produces No Viable Edge

The low valid trade count is not a surprise once the pre-sweep finding is understood. The AMD manipulation pattern fires on nearly half of all trading days, which means the signal itself is not rare. What is rare is a day where the signal fires, the target is not already reached before 9:40 AM, and the geometry still supports a minimum 1.5 R:R entry. Those conditions occur together about once every six weeks over this sample. That frequency is inconsistent with how practitioners describe their activity, which suggests the discretionary AMD trader is either applying looser versions of these rules, skipping some conditions based on contextual judgment, or working with a different conception of what constitutes a valid setup than the mechanical rules capture.

### 5.2 The Signal Predicts the Trend, Not a Reversal

Within this operationalization of the AMD model, no directional information beyond the daily trend regime was observed in the manipulation signal. Trend-aligned signals are correct 62–71% of the time; trend-opposed signals are correct only 25–30%. The effect sizes are medium-to-large (Cohen's $h = 0.23–0.53$) and all four trending-regime cells survive Benjamini-Hochberg correction. This is not the same as saying the signal has predictive value — it is saying the signal reflects the trend direction, and the trend is doing the predicting.

This finding is consistent with the sensitivity analysis: the inversion holds across manipulation thresholds from 5 to 20 NQ points, both close-back rule specifications, and EMA periods from 10 to 50. The consistency across specifications reduces, though does not eliminate, concerns that the result reflects an arbitrary threshold choice.

One result stands out for its practical implications. After London sweeps the Asia high in a bull market, price closes above the midpoint 75.2% of the time. The question of whether this inverted signal constitutes tradeable alpha after transaction costs cannot be answered from the current framework, which measures directional prediction at the daily level without defining a specific entry and exit mechanism. This is the central research question of Paper 2 in this series.

The regime dependency finding is consistent with what market microstructure theory would predict. In a sustained uptrend, institutional participants are net buyers over intermediate horizons. A London sweep of the Asia high in that environment may represent genuine accumulation above a resistance level — institutions filling buy orders just outside the range before continuing higher — rather than the deliberate false breakout the AMD model describes. This reading is consistent with Kyle's (1985) model of informed trading, in which large buyers push price upward to complete a position rather than engineer a reversal. An alternative interpretation, of course, is that the pattern reflects nothing institutional at all: in a bull market, any intraday dip that recovers will tend to produce an up close regardless of its cause, and the London sweep may simply be one variety of such a dip. We cannot distinguish between these explanations from bar-level OHLCV data alone; doing so would require order flow or tape data not available in this dataset.

In downtrend conditions, the mechanism may reverse: a London sweep of the Asia high in a bear market is more consistent with distribution before continuation lower, which aligns with what the AMD model predicts. The 59% bearish signal accuracy in 2022 during the Federal Reserve hiking cycle is at least consistent with this interpretation, though a sample of 61 bearish signal days in a single downtrend year is not sufficient to treat it as confirmation. One cannot rule out that 2022 was an outlier whose high accuracy reflects something other than the AMD mechanism — a year of unusually directional price action in which most signals happened to align with a dominant trend regardless of their cause.

### 5.3 Implications for ICT Practitioners and Researchers

These findings do not claim ICT practitioners are wrong to use the AMD model. A skilled discretionary trader reads higher-timeframe context, filters for setup quality, and exercises judgment no mechanical rule set can replicate. The mechanical rules tested here represent a lower bound — performance of the pattern without skill applied on top. Huddleston's documented public trading losses, including the 2016 challenge and 2024 Robbins Cup, suggest the model does not produce edge unconditionally even in expert hands, which is consistent with the regime dependency documented here.

For researchers, this study contributes a methodological template: a systematic approach to operationalizing a discretionary practitioner methodology into testable mechanical rules, with full documentation of threshold choices and comprehensive sensitivity analysis. The same approach is extensible to other ICT concepts including order blocks, fair value gaps, and the ICT market maker model.

### 5.4 Limitations and Future Work

Several limitations affect how strongly the findings should be interpreted. The macro exclusion calendar covers FOMC and NFP only; CPI and GDP release dates were not available from automated sources. Manual verification confirmed none of the six backtest trade dates fell on CPI or GDP releases, but any larger-sample extension of this work would need a complete exclusion calendar. The directional prediction analysis uses the daily close as the outcome measure, which is a fairly coarse proxy for whether the AMD signal was directionally correct; a finer-grained measure, such as the maximum favorable excursion during the New York session, might tell a different story. We also cannot rule out that the regime classification itself is driving some of the result: if uptrend and downtrend days differ systematically in ways beyond trend direction, those differences could partly explain the signal accuracy pattern without implicating the AMD mechanism at all. The six-trade backtest sample supports no formal inference, and a looser rule specification — requiring, say, six of nine conditions rather than all nine — might recover enough trades to run the statistical suite properly. How the results would look under such a specification is an open question.

Paper 2 in this series will test whether the AMD manipulation signal adds marginal directional predictive value above a simple trend-following baseline, and whether the inverted bearish uptrend signal constitutes tradeable alpha after transaction costs. Paper 3 will test an early-entry AMD variant that captures the distribution leg at or near the London close, directly addressing the timing incompatibility identified here as the primary source of setup attrition.

## 6. Conclusion

This paper set out to test whether the ICT AMD model produces statistically meaningful edge on CME E-Mini NQ futures when implemented as a strict mechanical strategy. The analysis spans N = 484 manipulation signal days across four years of real exchange data. Four findings emerge.

The AMD manipulation signal fires on 44% of trading days. On 57% of those days, price reaches the target before the 9:40 AM entry window opens — a structural property consistent across all calendar years and market regimes. The mechanical strategy produces six valid trades over four years, consistent with the timing finding and insufficient for formal backtest inference. Within this operationalization of the AMD model, no directional information beyond the daily trend regime was observed: trend-aligned signals are correct 62–71% of the time; trend-opposed signals are correct only 25–30%. All four trending-regime cells survive Benjamini-Hochberg correction with medium-to-large effect sizes, and the finding is robust to variation in the manipulation threshold, close-back rule specification, Asia range filter, and EMA period used for regime classification.

Two follow-on papers are motivated directly by these results. Paper 2 tests whether the AMD signal adds marginal value above trend following and whether the inverted bearish uptrend signal constitutes tradeable alpha. Paper 3 tests an early-entry variant designed to capture the distribution leg before the conventional New York window.

Beyond the specific findings, this paper provides a replicable methodological framework for quantitative evaluation of practitioner trading models. Full code and documentation are available at https://github.com/veertaylor19/amd-backtest. As ICT methodology continues to attract large numbers of retail traders, systematic empirical evaluation of its component models serves both the trading community and the academic literature on market microstructure and retail trader behavior.

## Appendix A: ICT Source Documentation

The following sources were consulted in deriving the mechanical rule set. Quantitative thresholds not present in any source represent the authors' operationalization decisions, identified throughout Section 3.

| Tier | Source | Relevance |
|---|---|---|
| Primary | Huddleston, M.J. [@ICTmentorship]. "2022 ICT Mentorship — Episode 9: Power Of Three & New York PM Session Opportunities." YouTube, 2022. https://www.youtube.com/playlist?list=PLVgHx4Z63paYiFGQ56PjTF1PGePL3r69s | Primary treatment of the Power of 3 / AMD concept: accumulation, manipulation, distribution phases; Judas Swing as the manipulation mechanism; New York session entry timing |
| Primary | Huddleston, M.J. [@ICTmentorship]. "2022 ICT Mentorship — Episode 10: Implementing Economic Calendar Events With The Open." YouTube, 2022. https://www.youtube.com/playlist?list=PLVgHx4Z63paYiFGQ56PjTF1PGePL3r69s | Power of 3 as Accumulation Manipulation Distribution; New York open price as reference; session structure for index futures |
| Secondary | innercircletrader.net/tutorials/ict-power-of-3/ (accessed May 2026) | Written AMD phase documentation, daily candle structure |
| Secondary | ttrades.com — ICT Power of Three AMD Explained (accessed 2026) | Session alignment, candle body and wick structure |
| Secondary | fxopen.com/blog/en/what-is-ict-po3-and-how-do-traders-use-it/ (accessed 2026) | Manipulation phase definition, close-back confirmation requirement |
| Secondary | innercircletrader.net/tutorials/master-ict-kill-zones/ (accessed 2026) | New York kill zone timing for index futures |

*Table A1. ICT practitioner sources consulted. Episodes 9 and 10 of the 2022 ICT Mentorship playlist are the primary sources for the AMD phase structure and session definitions used in this study. Both are available on the @ICTmentorship YouTube channel within the 2022 Mentorship playlist.*